\documentclass{jetpl}
\twocolumn

\title{Surface Instability of a Multi-Component Condensate and Andreev-Bashkin Effect}

\rtitle{Surface Instability ...}

\sodtitle{Surface Instability of a Multi-Component Condensate and Andreev-Bashkin Effect}

\author{D.\,A.\,Abanin\/\thanks{e-mail: abanin@itp.ac.ru}}

\rauthor{Abanin D.A.}

\sodauthor{Abanin}

\address{Low Temperature Laboratory, Helsinki University of Technology, P.O.Box 2200, FIN-02015, HUT, Finland\\~\\
Landau Institute for Theoretical Physics RAS, Kosygina 2, 117334 Moscow, Russia}

\dates{17 January 2003}{*}

\abstract{It is shown that surface of a liquid consisting of several interpenetrating superfluids moving with different velocities 
becomes unstable at some threshold. We demonstrate that the criterion for the onset of the instability changes in the presence of disspative interaction between the surface and the environment. Possible physical applications of the surface instability are discussed.}

\begin{document}

\maketitle

{\bf 1. Introduction.} 
Analogs of the classical Kelvin-Helmholtz instability are relevant for various physical systems. 
For instance, recently it was proposed [1] that the analog of KH instability for a mixture of two superfluids can
provide a possible mechanism for pulsar glitches [2]. The interest to KH instabilities has also been revived in connection with 
the observation of the instability arising on the interface separating two superfluids, $^3$He-A and $^3$He-B [3]. 
The results of this experiment are in excellent agreement with the criterion for the onset of the instability derived in [4]. 
Free surface instability of this class was discussed in [5]. 

However, in most interesting physical situations, such as neutron stars and spin Bose-Einstein condensates [6], several interpenetrating 
superfluids are present. Such mixtures of superfluids are known to exhibit Andreev-Bashkin effect [7] --- that is, the mass current 
of each condensate is carried by the velocities of the others:
\begin{equation}
{\bf j}_i=\sum_{k}\rho_{ik}{\bf v}^k, \;\;i=1,\dots,n, 
\end{equation}
where ${\bf v}^i$ and ${\bf j}_i$ are the superfluid velocity and the mass current of the $i$-th component of the condensate.    
 The denstiy of kinetic energy of such condensates can be written in the following form:
\begin{equation}
E_k=\frac 12\sum_{i,j}\rho_{ij} {\bf v}^i {\bf v}^j.
\end{equation}
 
In the present Letter we show that the free surface of a multi-component condensate exhibiting Andreev-Bashkin effect 
becomes unstable when velocities of its constituents reach certain values. There exist two different thresholds characterizing the instability, the first resulting from the dissipative interaction of the surface with the environment, which is typically small, and the second
corresponding to the classical dynamic instability. When the system is between these thresholds, the time of instability development 
is determined by the interaction with the enviroment; after passing through the second threshold the characteristic time of instability evolution quickly decreases. In some cases it is possible that there would be no intermediate region, that is, 
the instability develops very quickly from its very beginning.      

We propose that this surface instability should arise earlier than the bulk instability described in [1]. Indeed, the critical 
value of velocity obtained in [1] is comparable with the sound velocity in the bulk, while the critical velocity for arising surface instability is determined by the field stabilizing the surface and therefore should be much smaller. 

There is another possible application of the surface instability arising in mixtures of several interpenetrating superfluids: this mechanism could be relevant for the instability of a Bose-Einstein condensate with a multi-component order parmeter (see [6] and references therein).   

The paper is organized as follows. First we derive the criterion for the thermodynamic instability, then the condition for the emergence of dynamic instability is obtained. We analyze how the presence of the interaction of the surface with the environment affects the instability criterion. After that we demonstrate that in the 'shallow water' limit the physics of  the instability remains the same, but can be described in terms of the effective ripplon metric.

\vspace{3pt}

{\bf 2. Thermodynamic instability.}
In this section we consider a thermodynamic instability arising on a free surface of 
several interpenetrating superfluids moving with different superfluid velocities ${\bf v}^i$. We suppose that the 
surface in equilibrium coincides with the plane $z=0$ and superfluid velocities ${\bf v}^i$ are equal to ${\bf u}^i$. 
As we will show in Section 3, this instability results from the interaction between the surface of the liquid and the environment.        
The instability emerges when the energy of static perturbations becomes negative in the 
container frame (or, equivalently, in the rest frame of the normal components, where normal velocity ${\bf v}_n=0$). 

Let us find the energy of the perturbation with wave vector ${\bf k}=(k_x, k_y)$. First of all, the 
deformation $\zeta=a\sin {\bf kr}$ of the surface causes the perturbation $\delta {\bf v}^i=
\nabla  \Phi ^i$ of the superfluid velocity fields ${\bf v}^i$. Therefore, the free energy 
functional for the perturbation is given by:
\begin{multline}
{\cal F}=\frac 12 \int \, dx\, dy\, [F\zeta^2+\sigma \left((\partial _x\zeta)^2 +(\partial_y\zeta)^2\right) ]+{}\\
{}+\frac 12 \int _{-\infty}^{\zeta}\, dz \int \, dx\, dy\, [\sum _{i,j}\rho _{ij} \delta {\bf v}^i
\delta {\bf v}^j ] \, ,
\end{multline}
where $F$ is the external field stabilizing the surface, e.g., gravitational field, and $\sigma$ is surface tension. 
Since the perturbation is stationary, the continuity equation for each component reduces to $\Delta \Phi^i=0$.
Hence $\Phi^i$ has the following form:
\begin{equation}
\Phi^i=A^i\exp kz \cos{\bf kr}.
\end{equation} 
We can express constants $A^i$ via $a$ using the boundary conditions 
\begin{equation}
{\bf u}^i \frac {\partial \zeta}{\partial {\bf r}}=\frac{\partial \Phi^i}{\partial z} .
\end{equation}
This gives $A^i=a ({\bf u}^i\hat{{\bf k}}) $, where $\hat{\bf k}=\frac{\bf k}{k}$. Substituting these values of $A^i$ into the free energy (3), one obtains the energy of the perturbation:

\begin{equation}
E(k)=\frac{a^2}{4}[F+k^2 \sigma-k\sum_{i,j}\rho_{ij}({\bf u}^i \hat{\bf k})({\bf u}^j \hat{\bf k})].
\end{equation}
Thus, the criterion for the onset of the thermodynamic instability is given by
\begin{equation}
\frac 12\max_{\hat{\bf k}}\sum_{i,j} \rho_{ij} ({\bf u}^i \hat{\bf k})({\bf u}^j \hat{{\bf k}})   =\sqrt{\sigma F}. 
\end{equation}

\vspace{3pt}
{\bf 3. Dynamic instability.}
Now we proceed to the investigation of the analog of the classical Kelvin-Helmholtz dynamical
instability for several interpenetrating superfluids. This instability emerges when
the frequency of some surface wave acquires positive imaginary part. 
It appears that the threshold for arising of the instability depends substantially on the presence of dissipation. 
Even very small interaction of the superfluid with the environment (for example, it can be friction between the surface of superfluid 
and the container walls or between the surface and normal component) moves the threshold to another value. 
Let us first suppose that the interaction of the surface with the enviroment is absent.      

The Euler equation in our case reads
\begin{equation}
\sum _i\frac{\partial j^i_{\beta}}{\partial t}+\sum _{i,j}\rho_{ij} v^i_{\alpha}\nabla_{\alpha}
v_{\beta}^j =-\nabla_{\beta} p-F_{\beta}.   
\end{equation} 
The solution of (8) corresponding to a small-amplitude surface wave with frequency $\omega$
and wave vector ${\bf k}=(k_x, k_y)$ can be chosen in the following form:
\begin{equation}
v^i_x=u^i_x+ik_x e^{i({\bf kr}-\omega t)} e^{kz} A^i,
\end{equation}
\begin{equation}
v^i_y=u^i_y+ ik_y e^{i({\bf kr}-\omega t)} e^{kz} A^i
\end{equation}
\begin{equation}
v^i_z=k e^{i({\bf kr}-\omega t)} e^{kz} A^i,
\end{equation}
\begin{multline}
p=-Fz+i e^{i({\bf kr}-\omega t)} e^{kz}\times{}\\
{}\times [\omega \sum_{i,j}\rho_{ij}A^i -
\sum_{i,j}\rho_{ij}({\bf u}^i {\bf k}) A^j],
\end{multline}
where $A^i$, $i=1,\dots, n$ are some constants. 
Apart from equation (8), our system must satisfy boundary conditions:
\begin{equation}
-p=\sigma \left(  \frac{\partial^2}{\partial x^2}+ \frac{\partial ^2}{\partial y^2} \right) \zeta,
\end{equation}
\begin{equation}
\delta v_z^i-{\bf u}^i \frac{\partial \zeta}{\partial {\bf r}}=\frac{\partial \zeta}{\partial t}.
\end{equation}
Using equation (13) we can express the surface deformation $\zeta$ in terms of constants 
$A^i$:
\begin{multline}
\zeta=i e^{i({\bf kr}-\omega t)} e^{kz}/(F+k^2 \sigma){}\times\\
{}\times
[\omega \sum_{i,j}\rho_{ij}A^i -
\sum_{i,j}\rho_{ij}({\bf u}^i {\bf k}) A^j]
\end{multline}  
Substituting (15) into (14) and making use of (11), we obtain a set of equations for 
constants $A^i$:  
\begin{multline}
k(F+k^2\sigma) A^i=(\omega-{\bf u}^i {\bf k} )
[\sum_{j,l} \rho_{jl}A^j (\omega-{\bf u}^l {\bf k})], {}\\
 i=1,\dots, n.   
\end{multline}  
It is convinient to rewrite equation (16) in the form
\begin{equation}
\sum _j\phi_{ij} A^j=0, \;\; i=1,\dots, n, 
\end{equation}
where 
\begin{equation}
\phi_{ij}=k(F+k^2 \sigma)\delta_{ij} - (\omega-{\bf u}^i {\bf k})\sum_l \rho_{jl}(\omega-{\bf u}^l {\bf k}).
\end{equation}
For the system (17) to have a non-trivial solution it is neccesary that 
\begin{equation}
\det \phi_{ij}=0. 
\end{equation} 
It can be shown (see {\bf Appendix A}) that the condition (19) is equivalent to the following equation:
\begin{equation}
k(F+k^2 \sigma)=\sum_{ij}\rho_{ij}(\omega-{\bf u}^i {\bf k})(\omega-{\bf u}^j {\bf k}).  
\end{equation}
The equation (20) determines the ripplon spectrum. The frequency of the ripplon with wave vector directed along $\hat{\bf k}$  
acquires imaginary part when 
\begin{equation}
\rho\sum_{i,j}\rho_{ij}({\bf u}^i\hat{\bf k} ) ({\bf u}^j \hat{\bf k}) -{\left(\sum_{i,j}\rho_{ij} ({\bf u}^i \hat{\bf k} )   \right)^2}
=2\rho\sqrt{F\sigma},
\end{equation}
where $\rho=\sum_{i,j}\rho_{ij}$. The equation (21) provides the criterion for the onset of the instability which is 
analogous to the classical Kelvin-Helmholtz instability. In terms of 'mean velocity':
\begin{equation}
{\bf v}=\frac{{\bf j}_{\rm tot}}{\rho}=\frac{\sum_{i,j}\rho_{ij}{\bf u}^i }{\rho}
\end{equation}
equation (21) becomes more transparent:
\begin{equation}
\sum_{i,j}\rho_{ij}({\bf u}^i \hat{\bf k})({\bf u}^j \hat{\bf k})-\rho({\bf v}\hat{\bf k})^2=2\sqrt{F\sigma}.
\end{equation}

As was pointed out in [4], the criterion (23) will change if one takes into account dissipative force arising when the surface 
is moving with respect to the container walls. This force can be written in the following form:
\begin{equation}
F_{\rm fr}=-\Gamma \partial_t \zeta.
\end{equation}
In the presence of friction, the equation (20) for the spectrum of ripplons modifies as follows
\begin{equation}
k(F+k^2\sigma)-i\Gamma \omega k=\sum_{ij}\rho_{ij}(\omega-{\bf u}^i {\bf k})(\omega-{\bf u}^j {\bf k}). 
\end{equation}
When the superfluid velocities reach critical values, imaginary part of frequency crosses zero. Then from equation (25) it follows that 
the real part of frequency must also cross zero. Consequently, the instability condition is 
\begin{equation}
\frac 12 \max_{\hat{\bf k}} \sum_{i,j}\rho_{ij} ({\bf u}^i \hat{\bf k}) ({\bf u}^j \hat{\bf k})=\sqrt{\sigma F}. 
\end{equation}
This equation coincides with the condition (7) for the thermodynamic instability. From equations (23), (26) it can be easily seen that thermodynamic instability always arises earlier than the classical dynamic instability. In the region between these two thresholds the instability develops rather slowly (because the time of instability development is proportional to the friction parameter $\Gamma$, which is usually relatively small), and after passing through the dynamic threshold it becomes quickly developing. 

Criteria (23), (26) can be represented in another form, which sometimes proves more suitable. In order to obtain this form, we note that instability occurs first for the ripplons with wave vector $k_0=\sqrt{F/\sigma}$. We further rewrite equation (25) with ${\bf k}=k_0\hat{\bf k}$ in terms of the effective ripplon metric: 
\begin{equation}
\begin{array}{cc}
g^{\mu\nu}\hat{k}_{\mu}\hat{k}_{\nu}=i\Gamma\frac{\omega}{k_0} ,\\
\hat{k}_{\mu}=(-\omega/k_0, \hat{k}_{\alpha}), \; \alpha=x,y. \\
\end{array}
\end{equation}
The effective ripplon metric in equation (27) is given by:
\begin{equation}
\begin{array}{cc}
g^{00}=-\rho, \; g^{0\alpha}=-{j_{\rm tot}}_{\alpha}, \\
g^{\alpha\beta}=2\sqrt{F\sigma}\delta^{\alpha\beta}-\sum_{i,j}\rho_{ij}u^i_{\alpha} u^j_{\beta}.\\ 
\end{array}
\end{equation}
It can be easily seen that criterion (23) for dynamic instability is equivalent to the condition 
\begin{equation}
\det g_{\mu\nu}=\infty,
\end{equation}
while criterion (26) for thermodynamic instability is equivalent to 
\begin{equation}
g_{00}=0.
\end{equation}

\vspace{3pt}
{\bf 4. Case of two interpenetrating superfluids.}
Now let us consider in more detail the case when only two interpenetrating superfluids are present. The equation (25) for the spectrum of ripplons reads:
\begin{multline}
\frac{F+k^2\sigma}{k}-i\Gamma\frac{\omega}{k}=\rho_{11}(\frac{\omega}{k}-{\bf u}^1\hat{\bf k})^2\\
+\rho_{22}(\frac{\omega}{k}-{\bf u}^2\hat{\bf k})^2
+2\rho_{12}(\frac{\omega}{k}-{\bf u}^1\hat{\bf k})(\frac{\omega}{k}-{\bf u}^2\hat{\bf k}).
\end{multline}   
It is convenient to rewrite equation (31) 
in terms of mean velocity (22) and relative velocity ${\bf u}={\bf u}^1-{\bf u}^2$:
\begin{equation}
\frac{F+k^2\sigma}{k}-i\Gamma\frac{\omega}{k}=\rho\left[ (\frac{\omega}{k}-{\bf v}\hat{\bf k})^2+\alpha ({\bf u}\hat{\bf k})^2 \right], 
\end{equation}
where $\alpha=(\rho_1 \rho_2-\rho_{12}^2)/\rho^2$. Thus, the frequency of the surface wave with wave vector ${\bf k}$ satisfies the following equation:
\begin{equation}
\frac{\omega}{k}={\bf v}\hat{\bf k}\pm\sqrt{\frac{F+k^2\sigma}{\rho k}-i\frac{\Gamma}{\rho}\frac{\omega}{k}-\alpha({\bf u}\hat{\bf k})^2}. 
\end{equation}
Using (33), one immediately obtains that the first instability threshold, corresponding to the onset of the thermodynamic instability,
is determined by the equation:
\begin{equation}
\frac{2\sqrt{F\sigma}}{\rho}=\max_{\hat{\bf k}}\left[ \alpha ({\bf u}{\hat{\bf k}})^2+ ({\bf v}{\hat{\bf k}})^2\right],
\end{equation} 
and the second instability threshold is given by
\begin{equation}
\frac{2\sqrt{F\sigma}}{\rho}=\alpha u^2.
\end{equation}
Note that in the absence of Andreev-Bashkin effect results (34), (35) coincide with those for the instability of the interface separating two one-component superfluids [4]. The only difference between these situations consist in the fact that the stabilizing field $F$ in the case of surface instability is proportional to the total density of the mixture, while in the case of interface instability it is proportional to the difference of the densities of the separated liquids. 

As we have already mentioned, the thermodynamic instability always occurs before the dynamic one. However, under certain circumstances these instabilities can arise simultaneously. Indeed, if ${\bf u}$ is perpendicular to ${\bf v}$ and $v\leq \sqrt{\alpha} u$, then equations (34) and (35) coincide. Therefore, the instability starts at $u^2=\frac{2\sqrt{F\sigma}}{\rho}$ and it is 'strong' in the sense that the time of its development quickly decreases as $u$ grows. This scenario is analogous to the violation of the 'cosmic censorship' principle in quantum liquids [8], when the effective ripplon metric acquires naked singularity. 

{\bf 5. Shallow water limit.} In the previous sections we supposed that the depth of the liquid $h$ is large, so that 
\begin{equation}
\sqrt{\frac{F}{\sigma}}h\gg 1.
\end{equation} 
In this section we demonstrate that in the limit of 'shallow water' (i.e. $\sqrt{F/\sigma}h\ll 1$) the mechanism of the instability remains the same. Indeed, the equation for the ripplon spectrum in this limit has the following form:
\begin{equation}
\sum_{i,j}\rho_{ij}(\omega-{\bf u}^i{\bf k})(\omega-{\bf u}^j{\bf k})=F h  k^2+\sigma h k^4-i\Gamma(k)\omega. 
\end{equation}
It can be rewritten in terms of the effective ripplon metrics:
\begin{equation}
g^{\mu\nu}k_{\mu}k_{\nu}=i\Gamma(k)\omega-\sigma hk^4, 
\end{equation}
\begin{equation}
k_{\mu}=(-\omega,k_{\alpha}), \; \alpha=x,y. 
\end{equation}
\begin{equation}
g^{00}=-\rho, \; g^{0\alpha}=-{j_{\rm tot}}_{\alpha}
\end{equation}
\begin{equation}
g^{\alpha\beta}= Fh\delta^{\alpha\beta}-\sum_{ij}\rho_{ij}u^i_{\alpha} u^j_{\beta}. 
\end{equation}
Note that in contrast to the metric (28) the metric (40), (41) is valid for all long-wavelength ripplons. 

Again, there would be two thresholds for the onset of the instability, which are completely analogous to (23), (26), the only difference being that the ripplons which are responsible for the instability have wave vectors close to zero. The thermodynamic instability threshold corresponds to the condition (30) $g_{00}=0$, which ensures the appearance of the horizon, as was recently proposed by Schutzhold and Unruh [9]; the condition for the onset of the dynamic instability is $\det g_{\mu\nu}=\infty$, in agreement with [8].  

\vspace{3pt}
{\bf 6. Conclusion.} To conclude, we have found that the free surface of several interpenetrating superfluids, 
moving with different velocities, becomes unstable at some threshold. The criterion for such instability 
contains the off-diagonal densities $\rho_{ij}$, characterising Andreev-Bashkin effect. This kind of instability could provide a triggering mechanism for pulsar glitches. It also probably takes place in atomic Bose-Einstein condensates. 

\vspace{3pt}
{\bf Acknowledgements.} I am grateful to G. E. Volovik for suggesting this problem and for many useful discussions. 
I would also like to thank Low Temperature Laboratory of Helsinki University of Technology, where this work was done, 
for kind hospitality. This work was supported in part by RFFI grant 02-02-16218.

\vspace{3pt}
{\bf Appendix A.}
Here we present a calculation of the determinant of the matrix (18). It proves useful to introduce the following notations: 
\begin{equation}\alpha^i=\omega-{\bf u}^i {\bf k},
\end{equation}
\begin{equation}
\beta^l=\sum_i \rho_{il}(\omega-{\bf u}^i {\bf k}),
\end{equation}
\begin{equation}
\gamma=k(F+k^2\sigma).
\end{equation}
Then matrix (18) takes the form
\begin{equation}
\phi=
\left( \begin{array}{cccc} 
\gamma-\alpha^1\beta^1 & -\alpha^1\beta^2 & ... &-\alpha^1\beta^n \\
-\alpha^2\beta^1   & \gamma-\alpha^2\beta^2& ...& -\alpha^2\beta^n \\
...& ... & ... & ...\\
-\alpha^n \beta^1 & -\alpha^n \beta^2 & ... & \gamma-\alpha^n\beta^n \\
\end{array}
\right)
\end{equation}
The determinant of matrix $\phi_{ij}$ can be written as follows:
\begin{multline}
\det \phi =\prod_{i=1}^{n} \alpha^i\times \\
\times \det \left( 
\begin{array}{cccc}
\gamma/\alpha^1-\beta^1 & -\beta^2 & ... & -\beta^n \\
-\beta^1 & \gamma/\alpha^2-\beta^2 & ... & -\beta^n \\
... & ... & ... & ..\\
-\beta^1 & -\beta^2 & ... & \gamma/\alpha^n-\beta^n \\  
\end{array}
\right)
\end{multline}
In order to calculalate the determinant of the matrix on the r.h.s. of (46),
 let us subtract its $(n-1)$th line from its $n$th line, its $(n-2)$th line from 
its $(n-1)$th line, etc. This gives 
\begin{multline}
D_n=\\
\det \left( 
\begin{array}{cccc}
\gamma/\alpha^1 -\beta^1 & -\beta^2 & ...& -\beta^n \\
-\gamma/\alpha^1 & \gamma/\alpha^2 & 0 & 0 \\
... & ... & ... & ... \\
0 & 0 &
 -\gamma/\alpha^{n-1} &  \gamma/\alpha^n\\  
\end{array}
\right)
\end{multline}
One can easily obtain a reccurent formula for $D_n$:
\begin{equation}
D_n = \gamma /\alpha^n D_{n-1} - \beta^n \prod_{i=1}^{n-1} \gamma/\alpha^i  
\end{equation}
Thus, 
\begin{equation}
\det \phi = \gamma^n - \gamma^{n-1}\sum_{i=1}^n \alpha^i \beta^i, 
\end{equation}
and it equals zero when 
\begin{equation}
\gamma=\sum_{i=1}^n \alpha^i \beta^i.
\end{equation}

\end{document}